\documentclass[a4paper, final, natbib]{svjour3}    

\usepackage[english]{babel}
\usepackage{hyperref}
\usepackage{graphicx}
\usepackage{amsmath, amsbsy, amsfonts}
\usepackage{multirow}
\usepackage{lscape}
\usepackage{float}
\usepackage{rotating}
\usepackage{showkeys}
\usepackage{xcolor}
\def\Ascr{\mathcal{A}}
\def\Bscr{\mathcal{B}}
\def\Cscr{\mathcal{C}}

\def\Escr{\mathcal{E}}

\def\Gscr{\mathcal{G}}
\def\Hscr{\mathcal{H}}

\def\Lscr{\mathcal{L}}

\def\Oscr{\mathcal{O}}

\def\Rscr{\mathcal{R}}
\def\Sscr{\mathcal{S}}
\def\Tscr{\mathcal{T}}

\def\avec{\mathbf{a}}

\def\evec{\mathbf{e}}

\def\kvec{\mathbf{k}}

\def\nvec{\mathbf{n}}

\def\rvec{\mathbf{r}}

\def\Lambdavec{\boldsymbol{\Lambda}}
\def\lambdavec{\boldsymbol{\lambda}}
\def\xivec{\boldsymbol{\xi}}
\def\etavec{\boldsymbol{\eta}}

\def\Lvec{\mathbf{L}}
\def\Ivec{\mathbf{I}}

\def\sigmavec{\boldsymbol{\sigma}}
\def\rhovec{\boldsymbol{\rho}}
\def\omegavec{\boldsymbol{\omega}}
\def\dispchi{\raise2pt\hbox{$\chi$}}
\def\scriptchi{\raise1pt\hbox{$\scriptstyle\chi$}}
\def\rtvec{\tilde\rvec}
\def\lie#1{\Lscr_{#1}}
\def\epsilon{\varepsilon}
\def\rho{\varrho}
\def\phi{\varphi}
\def\csi{\xi}
\def\chr{{\it X$\rho$\'o$\nu o\varsigma$}}
\def\build#1_#2^#3{\mathrel{
\mathop{\kern 0pt#1}\limits_{#2}^{#3}}}

\begin{document}
\journalname{Celestial Mechanics and Dynamical Astronomy}

\title{Resonant Laplace-Lagrange theory for extrasolar systems in mean-motion resonance}

\author{M. Sansottera \and A.-S. Libert}

\institute{M. Sansottera\at
  Department of Mathematics, University of Milan, via Saldini 50, 20133 Milan, Italy\\
  \email{marco.sansottera@unimi.it}
  \and A.-S. Libert \at
  naXys, Department of Mathematics, University of Namur,
  8 Rempart de la Vierge,\\
  5000~Namur, Belgium \\
  \email{anne-sophie.libert@unamur.be}\\ 
  }        
\maketitle

\begin{abstract}
  Extrasolar systems with planets on eccentric orbits close to or in
  mean-motion resonances are common. The classical low-order resonant
  Hamiltonian expansion is unfit to describe the long-term evolution
  of these systems. We extend the Laplace-Lagrange secular
  approximation for coplanar systems with two planets by including
  (near-)resonant harmonics, and realize an expansion at high order in
  the eccentricities of the resonant Hamiltonian both at orders one
  and two in the masses. We show that the expansion at first order in
  the masses gives a qualitative good approximation of the dynamics of
  resonant extrasolar systems with moderate eccentricities, while the
  second order is needed to reproduce more accurately their orbital
  evolutions. The resonant approach is also required to correct the
  secular frequencies of the motion given by the Laplace-Lagrange
  secular theory in the vicinity of a mean-motion resonance. The
  dynamical evolutions of four (near-)resonant extrasolar systems are
  discussed, namely GJ~876 (2:1 resonance), HD~60532 (3:1), HD~108874
  and GJ~3293 (close to 4:1).

\end{abstract}

\section{Introduction}\label{sec:intro}
The search for exoplanets around nearby stars has produced a
tremendous amount of observational data, pointing out the peculiar
character of the Solar System. To date, more than 600
  multiple planet systems have been found and the number of discovered
  planets with unexpected orbital properties (such as highly eccentric
  orbits, mutually inclined planetary orbits, hot Jupiters, compact
  multiple systems...) constantly increases. Extrasolar systems that
reside in or near mean-motion resonances are commonly detected with
significantly high eccentricities. Therefore, the classical approach
based on low-order expansions in eccentricities conceived
for the Solar System is at least questionable when applied to
extrasolar systems.

In the non-resonant scenario, the long-term evolution of a planetary
system is described by the secular theory which consists in an
averaging of the Hamiltonian over the fast angles (related to the mean
anomalies). The classical procedure, denoted ``averaging by
scissors'', removes from the Hamiltonian all the terms including the
fast angles.  Thus the fast actions are fixed, and so are the
semi-major axes.  This Hamiltonian is named approximation at {\it
  order one in the masses} and the linearized equations of motion
correspond to the Laplace-Lagrange secular theory.  This approach is
adequate for the quasi-circular planetary orbits of the Solar System.
Pushing the expansions at higher order terms in the eccentricities (up
to order 12) in~\cite{LibHen-2005,LibHen-2007}, the authors have shown
that the high-order expansion accurately reproduces the long-term
dynamics of extrasolar systems with eccentric orbits and which are far
away from mean-motion resonance.

Instead, if the ratio between the mean-motion frequencies is close to
the ${k_1^*}:{k_2^*}$ commensurability, the impact of the harmonics
$(k_1^* \lambda_1 -k_2^* \lambda_2)$ on the long-term evolution of the
system should be included in the Hamiltonian formulation.  Still, in
view of the exponential decay of the Fourier expansion with
$|\kvec|_1=|k_1|+|k_2|$, only low-order mean-motion resonances must be
considered. The effect of near-resonances on the long-term evolution
of a planetary system is taken into account by considering the
so-called Hamiltonian at {\it order two in the masses}, where the
integrable approximation describes an invariant torus (up to order two
in the masses) instead of the classical circular orbits. The gain
obtained with this approximation in the context of the planetary
motion of the Solar System has been deeply studied
in~\cite{Laskar-1988}. Also, the second order approximation plays a
crucial role in the applicability of the celebrated theorems of
Kolmogorov and Nekhoroshev to the planetary case since it allows to
tackle analytically the problem of the stability of the Solar System,
see e.g.,~\cite{Robutel-1995}, \cite{CelChi-2005},
\cite{GabJorLoc-2005}, \cite{LocGio-2007}, \cite{GioLocSan-2009,
  GioLocSan-2017}, \cite{SanLocGio-2011a, SanLocGio-2011b}, and
\cite{SanGraGio-2015}. Regarding extrasolar systems,
in~\cite{LibSan-2013}, we have considered a secular Hamiltonian at
order two in the masses and showed that this secular Hamiltonian
provides a good approximation of the dynamics for extrasolar systems
which are {\it near a mean-motion resonance}, but not {\it really
  close to} or {\it in a mean-motion resonance}. Furthermore, the
canonical transformation used for the approximation at second order in
the masses allowed to evaluate the proximity of the system to a
mean-motion resonance.  In particular, we defined an heuristic
criterion based on the computation of a $\delta$-parameter which
allows to discriminate between the different regimes (see
Sect.~\ref{ss:delta} and also~\cite{LibSan-2013} for more details).

Coming to the resonant case, several alternatives have been proposed
to study analytically the dynamics of resonant extrasolar
systems. Instead of expanding the disturbing function in series of the
eccentricities following the classical approach, \cite{Bea-2003} have
introduced an analytic expansion based on a linear regression, which
is convenient for the high-eccentricity planetary three-body
problem. Other works (e.g., \cite{Alv-2015}) avoid the use of
expansions and consider the adiabatic regime, eliminating all the
short-period terms by directly performing a numerical averaging of
the Hamiltonian. For systems with small planetary eccentricities, the
classical low-order expansion in eccentricities is suitable, as
highlighted for instance in~\cite{Cal-2004} for the near-resonance
between Uranus and Neptune and in~\cite{Cal-2006} for a 3:2 resonance
among planets around the pulsar PSRB1257+12.  The limitations of an
expansion at first order in the masses and truncated at order four in
the eccentricities have been studied in~\cite{Ver-2007}, making
reference to the GJ~876 extrasolar system.  Finally, an integrable
approximation for first order mean-motion resonance (i.e., a $k:k-1$
resonance) has been developed in~\cite{Bat-2013} in order to address
resonant encounters during divergent migration.

In the present work we are interested in the dynamics of systems which
are really close to or in a mean-motion resonance. Our strategy is to
reconstruct their evolution using a resonant Hamiltonian expanded at
high orders, hence including appropriate resonant combinations of the
fast angles.  In other words, our goal is to extend the classical
Laplace-Lagrange secular theory to the resonant case.  We aim to see
whether it is worth pushing the expansion of the resonant Hamiltonian
to high order in eccentricities and to the order two in the masses for
describing the evolutions of resonant exoplanets with moderate
eccentricity.

The problem tackled here is challenging: the convergence domain of the
Laplace-Lagrange expansion of the disturbing function is limited, as
demonstrated by~\cite{Sun-1916} (see also~\cite{FM-1994}). Considering
only the secular terms of the expansion,
\cite{LibHen-2005,LibHen-2007} observed a numerical convergence of the
secular expansion ({\it convergence au sens des astronomes}, see
\cite{Poi-1893}) and showed that the approximation is accurate even
for moderate to high planetary eccentricities.  However, here we must
consider also the contribution of the resonant combinations, making
this approach potentially doubtful for extrasolar systems.  As a
result, we think that establishing the extent of validity of the
classical approaches, along the lines of classical Celestial
Mechanics, is pivotal and deserves to be investigated.

The paper is organized as follows. In Section~\ref{sec:expansion}, we
introduce the planar Poincar\'e variables and outline the expansion of
the Hamiltonian.  The resonant Hamiltonians, both at orders one and
two in the masses, are presented in Section~\ref{sec:resonant}, while
Section~\ref{sec:appl} is dedicated to the application of our
analytical approach to several (near-)resonant extrasolar systems,
where representative planes of the dynamics and the proximity to
periodic orbits family are also discussed. Finally, our findings are
summarized in Section~\ref{sec:ccl}.

\section{Hamiltonian expansion}\label{sec:expansion}

We consider a system consisting of a central star of mass $m_0$ and
two coplanar planets of masses $m_1$ and $m_2$. The indices $1$ and
$2$ refer to the inner and outer planets, respectively. The
Hamiltonian formulation of the planetary three-body problem in
canonical heliocentric variables (see, e.g.,~\cite{Laskar-1989}) with
coordinates $\rvec_j$ and momenta $\tilde{\rvec}_j$, for $j = 1,\, 2$,
has four degrees of freedom, and reads
\begin{equation}
F(\rvec,\rtvec)=
T^{(0)}(\rtvec)+U^{(0)}(\rvec)+
T^{(1)}(\rtvec)+U^{(1)}(\rvec)\ ,
\label{eq:H_iniz}
\end{equation}
where
\begin{alignat*}{2}
T^{(0)}(\tilde{\bf r})&=\frac{1}{2}\sum_{j=1}^2 \|\tilde{\bf r}_j\|^2 \left(\frac{1}{m_0}+\frac{1}{m_j} \right)\ , \qquad&
T^{(1)}(\tilde{\bf r})&= \frac{\tilde{\bf r}_1 \cdot \tilde{\bf r}_2}{m_0}\ ,\\
U^{(0)}({\bf r})&=-\Gscr\sum_{j=1}^2 \frac{m_0m_j}{\|{\bf r}_j\|}\ , \qquad&
U^{(1)}({\bf r})&= -\Gscr\frac{m_1m_2}{\|{\bf r}_1-{\bf r}_2\|}\ .
\end{alignat*}
When adopting the planar Poincar\'e canonical variables, i.e.,
\begin{equation}
\vcenter{\openup1\jot\halign{
 \hbox {\hfil $\displaystyle {#}$}
&\hbox {\hfil $\displaystyle {#}$\hfil}
&\hbox {$\displaystyle {#}$\hfil}
&\hbox to 6 ex{\hfil$\displaystyle {#}$\hfil}
&\hbox {\hfil $\displaystyle {#}$}
&\hbox {\hfil $\displaystyle {#}$\hfil}
&\hbox {$\displaystyle {#}$\hfil}\cr
\Lambda_j &=& \frac{m_0\, m_j}{m_0+m_j}\sqrt{\vphantom{b^a}\Gscr(m_0+m_j) a_j}\ ,
& &\lambda_j &=& M_j+\omega_j\ ,
\cr
\csi_j &=& \sqrt{\vphantom{b^a} 2\Lambda_j}\,
\sqrt{1-\sqrt{1-e_j^2}}\,\cos\omega_j\ ,
& &\eta_j&=&-\sqrt{\vphantom{b^a}2\Lambda_j}\,
\sqrt{1-\sqrt{1-e_j^2}}\, \sin\omega_j\ ,
\cr
}}
\label{eq:poincvar}
\end{equation}
for $j=1\,,\,2\,$, where $a_j\,,\> e_j\,,\> M_j$ and $\omega_j$ are
the semi-major axis, the eccentricity, the mean anomaly and the
longitude of the pericenter of the $j$-th planet, respectively, we
have
\begin{equation}
F(\Lambdavec, \lambdavec, \xivec, \etavec)=
F^{(0)}(\Lambdavec)+F^{(1)}(\Lambdavec, \lambdavec, \xivec
,\etavec)\ ,
\label{eq:H_iniz_poinc}
\end{equation}
where $F^{(0)}=T^{(0)}+U^{(0)}$ is the Keplerian part and
$F^{(1)}=T^{(1)}+U^{(1)}$ is the perturbation (see, e.g.,~\cite{Laskar-1989}). The ratio between the two parts is of order $\Oscr(\mu)$ with $\mu=\max\{m_1/m_0,m_2/m_0\}$, so the variables $(\Lambdavec,\lambdavec)$ are referred to as the fast variables and $(\xivec,\etavec)$ as the secular variables. Following \cite{LibSan-2013}, we realize an expansion of the Hamiltonian in Taylor-Fourier series of the Poincar\'e variables.

To do so, we first realize a translation, $\Tscr_{F}$, in the fast actions
$$
\Lvec=\Lambdavec-\Lambdavec^*\ , 
$$
where $\Lambdavec^*$ is a fixed value\footnote{Here we expand around
  the initial values, but the average values over a long-term
  numerical integration could also be considered (see,
  e.g.,~\cite{SanLocGio-2011a}).}. The
Hamiltonian~\eqref{eq:H_iniz_poinc} is then expanded in Taylor series
of $\Lvec$, $\xivec$ and $\etavec$ around the origin, as well as in
Fourier series of $\lambdavec$, and we obtain
\begin{equation}
\Hscr^{(\Tscr)}=\nvec^*\cdot\Lvec+\sum_{j_1=2}^{\infty} h_{j_1,0}^{{\rm (Kep)}}(\Lvec)+
\mu\sum_{j_1=0}^{\infty}\sum_{j_2=0}^{\infty} h^{(\Tscr)}_{j_1,j_2}(\Lvec,\lambdavec,\xivec,\etavec)\ ,
\label{eq:H_trasl}
\end{equation}
with $n_j^* = \sqrt{(m_0+m_j)/a_j^3}$ for $j=1,2\,$, and where the
terms $h_{j_1,0}^{{\rm (Kep)}}$ are homogeneous polynomials of degree
$j_1$ in the fast actions~$\Lvec$, while the functions
$h^{(\Tscr)}_{j_1,j_2}$ are homogeneous polynomials of degree $j_1$ in
the fast actions~$\Lvec$, degree $j_2$ in the secular variables
$(\xivec,\etavec)$, and trigonometric polynomials in the
angles~$\lambdavec$. A detailed treatment can be found
in~\cite{Duriez-1989a, Duriez-1989b, Laskar-1989}
and~\cite{LasRob-1995} where the low-order analytical expressions of
the crucial terms are given in Section~7. The computations have been
done via algebraic manipulation, using a package developed on purpose
named {\chr} (see~\cite{GioSan-2012}).

The expansion is truncated as follows. We include in the Keplerian
part terms up to degree $2$ in the fast actions $\Lvec$, while in the
perturbation we consider the terms which are: (i)~up to degree $1$ in
$\Lvec$; (ii)~up to degree $12$ in the secular variables
$(\xivec,\etavec)$; (iii)~up to trigonometric degree $24$ in the fast
angles $\lambdavec$.

\section{Resonant approximation of the Hamiltonian}\label{sec:resonant}

In the present work we are interested in the dynamics of systems that
are really close to or in a mean-motion resonance. The averaging
procedures, both at first and second orders in the masses, must be
adapted so as to take into account the strong influence of the
mean-motion resonance.  The main issue concerns the presence of small
divisors, that prevents the convergence of the averaging procedure on
a domain that contains the initial data.  We explore here two
different approaches, considering {\it resonant} Hamiltonians at
order one and two in the masses.

\subsection{Resonant Hamiltonian at order one in the masses}\label{sbs:resO1}
The classical ``averaging by scissors'' method can be simply modified
by including appropriate resonant combinations of the fast angles into
the Laplace-Lagrange secular expansion. The resonant combination to
take into account is related to the nearest mean-motion resonance and
can be determined from the vector $\nvec^*$, for example by means of
continued fraction approximations, or by exploiting the
$\delta$-parameter (see Sect.~\ref{ss:delta}).  Precisely,
having fixed a resonant harmonic $\kvec^*\cdot\lambdavec$, the
resonant Hamiltonian at order one in the masses writes
\begin{equation}
\Hscr_{{\rm res}}^{(\Oscr 1)} = \overline\Hscr^{(\Tscr)} + \widetilde\Hscr^{(\Tscr)}\ ,
\label{eq:HresO1}
\end{equation}
with
$$
\overline\Hscr^{(\Tscr)} = \frac{1}{4\pi^2}
\int_{0}^{2\pi}\!\!\!\int_{0}^{2\pi} \Hscr^{(\Tscr)}
\,{\rm d}\lambda_1\,{\rm d}\lambda_2
\quad\textrm{and}\quad
\widetilde\Hscr^{(\Tscr)} = \mu\sum_{j_1=0}^{\infty}\sum_{j_2=0}^{\infty} h^{(\Tscr)}_{j_1,j_2}(\Lvec,\kvec^*\cdot\lambdavec,\xivec,\etavec)\ .
$$
The first term, $\overline\Hscr^{(\Tscr)}$, is the classical averaging
of the Hamiltonian over the fast angles, $\lambdavec$.  In the second
one, $\widetilde\Hscr^{(\Tscr)}$, all the terms corresponding to the
resonant harmonic $\kvec^*\cdot\lambdavec$ and its multiples are
collected. Although they depend on the fast angles, due to the
(near-)resonance relation, they are {\sl slow} terms and have to be
taken into account in the study of the long-term evolution.

\subsection{Resonant Hamiltonian at order two in the masses}\label{sbs:resO2}
The secular approximation at order two in the masses is based on a
``Kolmogorov-like'' normalization step aiming at removing the fast
angles from terms that are at most linear in the fast actions~$\Lvec$.
Again, we modify the standard approach by putting the resonant
combinations of the fast angles, the harmonics
$\kvec^*\cdot\lambdavec$, in the normal form, thus removing them from
the generating function.

We briefly summarize the averaging procedure here, more details can be
found in, e.g.,~\cite{LocGio-2007, SanLocGio-2011a} and
\cite{LibSan-2013}.  We adopt the standard Lie series algorithm (see,
e.g.,~\cite{Henrard-1973} and \cite{Giorgilli-1995}) to transform the
Hamiltonian~\eqref{eq:H_trasl} into $\widehat\Hscr^{(\Oscr 2)}=\exp
\lie{\mu\,\scriptchi_{1}^{(\Oscr 2)}}\,\Hscr^{(\Tscr)}$.  The
generating function $\mu\, \dispchi_1^{(\Oscr
  2)}(\lambdavec,\xivec,\etavec)$ is determined by solving the
homological equation
\begin{equation}
\sum_{j=1}^{2}n^*_j \frac{\partial\,\dispchi_{1}^{(\mathcal{O} 2)}}{\partial \lambda_j}
+\sum_{j_2=0}^{K_S}\left\lceil h_{0,j_2}^{(\Tscr)} \right\rceil_{\lambdavec;K_F}
(\lambdavec,\xivec,\etavec)=\sum_{j_2=0}^{K_S}\left\lceil h_{0,j_2}^{(\Tscr)} \right\rceil_{\lambdavec;K_F}
(\kvec^*\cdot\lambdavec,\xivec,\etavec)\ ,
\label{eq:chi_1}
\end{equation}
where $\lceil f \rceil_{\lambdavec;K_F}$ denotes the Fourier expansion
of a function $f$ including only the harmonics satisfying
$0<|\kvec|_1\leq K_F$.  The parameters $K_S$ and $K_F$ are tailored
according to the considered mean-motion resonance, the actual choice
is detailed in Section~\ref{sec:appl} for each system studied
here. The transformed Hamiltonian $\widehat {\mathcal H}^{(\mathcal{O}
  2)}$ can be written in the same form as~\eqref{eq:H_trasl},
replacing $h^{(\Tscr)}_{j_1,j_2}$ with $\hat h^{(\mathcal{O}
  2)}_{j_1,j_2}$. Here and in the following, with a common little
abuse of notation, we denote the new variables with the same names as
the old ones in order to not unnecessarily burden the notation.

To compute the resonant Hamiltonian to the second order in the masses,
$\Hscr^{(\Oscr 2)}=\exp \lie{\mu\,\scriptchi_{2}^{(\Oscr
    2)}}\,\widehat\Hscr^{(\Oscr 2)}$, we solve the following
homological equation to determine the generating function
$\mu\,\dispchi_{2}^{(\Oscr 2)}(\Lvec,\lambdavec,\xivec,\etavec)$,
which is linear in $\Lvec$,
\begin{equation}
  \sum_{j=1}^{2}n^*_j \frac{\partial\,\dispchi_{2}^{(\Oscr
      2)}}{\partial \lambda_j} +\sum_{j_2=0}^{K_S}\left\lceil
  \hat{h}_{1,j_2}^{(\Oscr 2)} \right\rceil_{\lambdavec;K_F}
  (\Lvec,\lambdavec,\xivec,\etavec)=\sum_{j_2=0}^{K_S}\left\lceil
  \hat{h}_{1,j_2}^{(\Oscr 2)} \right\rceil_{\lambdavec;K_F}
  (\Lvec,\kvec^*\cdot\lambdavec,\xivec,\etavec)\ .
  \label{eq:chi_2}
\end{equation}
The resonant Hamiltonian $\mathcal{H}^{(\mathcal{O} 2)}$ writes
\begin{equation}\
  \Hscr^{(\Oscr 2)}(\Lvec,\lambdavec,\xivec,\etavec)=
  \nvec^*\cdot\Lvec+
  \sum_{j_1=2}^\infty h_{j_1,0}^{({\rm Kep})}(\Lvec)+
  \mu\sum_{j_1=0}^\infty\sum_{j_2=0}^\infty
  h_{j_1,j_2}^{(\Oscr 2)}(\Lvec,\lambdavec,\xivec,\etavec;\mu)+\Oscr(\mu^3)\ .
  \label{eq:H_ord2}
\end{equation}
Note that we denote again by $(\Lvec,\lambdavec,\xivec,\etavec)$ the
new coordinates.  The composition of the Lie series with generating
functions $\mu\,\dispchi_{1}^{(\Oscr 2)}$ and
$\mu\,\dispchi_{2}^{(\Oscr 2)}$, will be denoted by $\Tscr_{\Oscr 2}$,
precisely
\begin{equation}
\Tscr_{\Oscr 2}(\Lvec,\lambdavec,\xivec,\etavec) = 
\exp \lie{\mu\,\scriptchi_{2}^{(\Oscr 2)}} \circ
\exp \lie{\mu\,\scriptchi_{1}^{(\Oscr 2)}}(\Lvec,\lambdavec,\xivec,\etavec)\ .
\label{eq:trasf_ord2}
\end{equation}

Finally, since we aim for a second order approximation, we neglect the
terms of order $\Oscr(\mu^3)$ and, likewise we do in the first order
approximation, we select the secular and resonant terms as
in~\eqref{eq:HresO1}, namely
\begin{equation}
\Hscr_{{\rm res}}^{(\Oscr 2)} = \overline\Hscr^{(\Oscr 2)} + \widetilde\Hscr^{(\Oscr 2)}\ .
\label{eq:HresO2}
\end{equation}
Similarly to the resonant approximation at order one in the masses,
$\overline\Hscr^{(\Oscr 2)}$ denotes the Hamiltonian $\Hscr^{(\Oscr
  2)}$ averaged over the fast angles, $\lambdavec$, while
$\widetilde\Hscr^{(\Oscr 2)}$ contains the terms corresponding to the
resonant harmonic $\kvec^*\cdot\lambdavec$ and its multiples. Let us
stress the crucial difference between the first and second order
approximations: the second order one takes into account the effect of
low-order harmonics in both expressions $\overline\Hscr^{(\Oscr 2)}$
and $\widetilde\Hscr^{(\Oscr 2)}$.
    
To illustrate the averaging procedure described here, the expressions
of the resonant Hamiltonians at order one and two in the masses are
given in the Appendix~\ref{sec:A_exp} for GJ~876 system (see
Subsection~\ref{sbs:GJ876} for a complete description of the
system). All the terms are reported up to degree $1$ in the fast
actions $\Lvec$, degree $6$ in the fast angles $\lambdavec$ and degree
$2$ in the secular variables $(\xivec, \etavec)$.

\subsection{the $\delta$-parameter: proximity to a mean-motion resonance}\label{ss:delta}
In~\cite{LibSan-2013} we introduced an heuristic criterion to evaluate
the proximity of planetary systems to mean-motion resonance, by
exploiting the canonical change of coordinates used for the
approximation at order two in the masses.  We report here the
definition of the $\delta$-parameter that will be used in the analysis
of the selected planetary systems, and refer to~\cite{LibSan-2013} for
a detailed discussion.

The first order terms of the near the identity change of variables,
$\Tscr_{\Oscr2}$, are
\begin{align*}
  \xi_j'&=\xi_j\,\left( 1-\frac{1}{\xi_j}\frac{\partial\,\mu\,\dispchi_1^{(\Oscr 2)}}{\partial\eta_j}\right)\ ,\\
  \eta_j'&=\eta_j\,\left( 1-\frac{1}{\eta_j}\frac{\partial\,\mu\,\dispchi_1^{(\Oscr 2)}}{\partial\xi_j}\right)\ ,
\end{align*}
for $j=1\,,\,2\,$.  Considering the coefficients of the functions
\begin{equation}
  \delta\xi_j = \frac{1}{\xi_j}\frac{\partial\,\mu\,\dispchi_1^{(\Oscr 2)}}{\partial\eta_j}
  \qquad\hbox{and}\qquad
  \delta\eta_j = \frac{1}{\eta_j}\frac{\partial\,\mu\,\dispchi_1^{(\Oscr 2)}}{\partial\xi_j}\ ,
  \label{eq:prox}
\end{equation}
we identify the most important (near) mean-motion resonance terms
corresponding to the harmonic $\kvec\cdot\lambdavec$.  Given a
polydisk of radius $\rhovec$ around the origin, $\Delta_{\rhovec}$,
and adopting the weighted Fourier norm, we define the quantities
$$
\delta\xi_j^* = \max_{\kvec}(\|\delta\xi^{(\kvec)}_j\|_{\rhovec})
\qquad\hbox{and}\qquad \delta\eta_j^* =
\max_{\kvec}(\|\delta\eta^{(\kvec)}_j\|_{\rhovec})
$$
needed for the computation of the $\delta$-parameter given by
$\delta=\max(\delta_1, \delta_2)\,$ where $\delta_j =
\min(\delta\xi_j^*, \delta\eta_j^*)$ for $j=1,2$.  The $\delta$-parameter
is a measure of the change from the original secular
variables to the averaged ones.  The actual computation of the $\delta$-parameter is
quite cumbersome, but is more reliable than just looking at the
semi-major axes ratio, since it holds information about the non-linear
character of the system.

In particular, in~\cite{LibSan-2013}, we defined an heuristic
criterion based on the computation of the $\delta$-parameter to
discriminate between three different regimes: (i)~if
$\delta<2.6\times10^{-3}$, the system is far from the mean-motion
resonance, so the first order secular approximation describes the
long-term evolution of the system with great accuracy; (ii)~if
$2.6\times10^{-3}<\delta\leq2.6\times10^{-2}$, a secular Hamiltonian
at order two in the masses is required to describe the long-term
evolution; (iii)~if $\delta>2.6\times10^{-2}$, the system is too close
to a mean-motion resonance and a secular approximation is not enough
to describe the long-term evolution.

\subsection{Resonant variables}\label{sbs:res}

The resonant Hamiltonians at order one and two in the masses,
$\Hscr_{{\rm res}}^{(\Oscr 1)}$ and $\Hscr_{{\rm res}}^{(\Oscr 2)}$,
respectively, reduce the problem to two degrees of freedom.  To
highlight this point, we introduce the canonical resonant variables
associated to a ${k_1^*}:{k_2^*}$ mean-motion resonance (see for
instance \cite{Bea-2003}).  The canonical change of coordinate
$\Tscr_\Rscr$ is given by
\begin{equation}
\vcenter{\openup1\jot\halign{
 \hbox {\hfil $\displaystyle {#}$}
&\hbox {\hfil $\displaystyle {#}$\hfil}
&\hbox {$\displaystyle {#}$\hfil}
&\hbox to 6 ex{\hfil$\displaystyle {#}$\hfil}
&\hbox {\hfil $\displaystyle {#}$}
&\hbox {\hfil $\displaystyle {#}$\hfil}
&\hbox {$\displaystyle {#}$\hfil}\cr
J_1=L_1+ \frac{k_2^*}{{k_1^*}-{k_2^*}} (I_1+I_2) \ , & & \lambda_1 \ ,
\cr
J_2=L_2- \frac{k_1^*}{{k_1^*}-{k_2^*}} (I_1+I_2) \ , \quad & & \lambda_2 \ ,
\cr
I_1= \sqrt{2\Lambda_1} \sqrt{1-\sqrt{1-e_1^2}} \ , & & \sigma_1=-\frac{k_2^*}{{k_1^*}-{k_2^*}}\lambda_1+\frac{k_1^*}{{k_1^*}-{k_2^*}}\lambda_2-\omega_1 \ ,
\cr
I_2= \sqrt{2\Lambda_2} \sqrt{1-\sqrt{1-e_2^2}} \ , & & \sigma_2=-\frac{k_2^*}{{k_1^*}-{k_2^*}}\lambda_1+\frac{k_1^*}{{k_1^*}-{k_2^*}}\lambda_2-\omega_2 \ ,\cr
}}
\label{eq:resonantvar}
\end{equation}
The resonant Hamiltonians~\eqref{eq:HresO1} and~\eqref{eq:HresO2}
contain two types of terms only: (i)~secular terms which have no
dependency in the fast angles; (ii)~resonant terms depending on the
fast angles through the resonant angles $\sigma_1$ and $\sigma_2$. As
a result, $J_1$ and $J_2$ are constants of motions and the resonant
Hamiltonians have only two degrees of freedom $(I_1,\sigma_1)$ and
$(I_2,\sigma_2)$. In the next section, we investigate the limitations
and/or improvements of the analytical expansion detailed here in
modeling the long-time dynamics of extrasolar systems really close to
or in a mean-motion resonance.

\section{Application to (near-)resonant extrasolar systems}\label{sec:appl}

\begin{figure}[tb]
  \begin{center}
    \includegraphics[width=\textwidth]{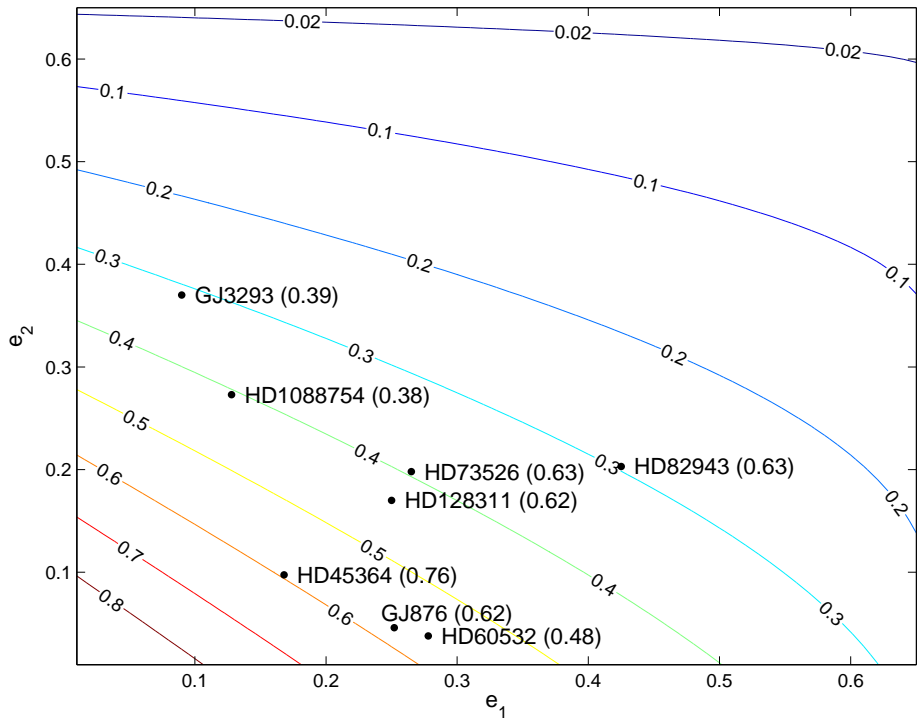}
  \end{center}
  \caption{Sundman's convergence domain for the selected extrasolar
    systems. The lines represent the boundary (in eccentricities)
    delimitation of the convergence domain (below area), for different
    semi-major axis ratios. Initial parameters of several extrasolar
    systems are shown, with the planetary semi-major axis ratio given
    in parenthesis.}
 \label{figsundman}
\end{figure}

We selected 8 extrasolar systems in the vicinity of a low-order
mean-motion resonance for which a full parameterization has been
derived, namely GJ~876 (\cite{Lau-2001}), HD~128311 (\cite{Vogt-2005}),
HD~73526 (\cite{Wit-2014}) and HD~82943 (\cite{Tan-2013}) for the 2:1
mean-motion resonance, HD~45364 (\cite{Cor-2009}) for the 3:2
resonance, HD~60532 (\cite{Laskar-2009}) for the 3:1 resonance, as
well as HD~108874 (\cite{WriUpaMarFis-2009}) and GJ~3293
(\cite{Ast-2015}) for the 4:1 resonance.

\begin{table}[tb]
\caption{Parameters of GJ~876 (\cite{Lau-2001}), HD~60532 (\cite{Laskar-2009}), HD~108874 (\cite{WriUpaMarFis-2009}) and GJ~3293 (\cite{Ast-2015}) extrasolar systems.}
\label{tab:param}
\begin{center}
\begin{tabular}{llllllllll}
\hline
System & $m$ ($M_{\rm Jup}$) & $a$ & $e$ & $\omega$ ($\deg$) & $M$ ($\deg$) & $\delta$ \\
\hline
GJ~876c & 0.92 & 0.1291& 0.252& 198.30& 308.80&\multirow{2}{*}{$1.6$} \\
GJ~876b & 3.08 & 0.2067& 0.046& 176.80& 174.30 & \\
HD~60532b& 3.1548 & 0.7606  & 0.278 & 352.83& 21.950 &\multirow{2}{*}{$6.8\times 10^{-1}$} \\
HD~60532c& 7.4634 & 1.5854 & 0.038 & 119.49& 197.53 & \\
HD~1008874b& 1.34 & 1.053& 0.128& 219.49& 0.0 &\multirow{2}{*}{$2.3\times 10^{-2}$}\\
HD~1008874c& 1.064& 2.77& 0.273& 10.0& 267.43 & \\
GJ~3293b& 0.0812& 0.1434 &0.09 & 282.3& 0.0& \multirow{2}{*}{$2.1\times 10^{-2}$} \\
GJ~3293c& 0.0705& 0.364& 0.37& 322.0&230.50 & \\
\hline
\end{tabular}
\end{center}
\end{table}

As previously explained, the convergence of the Laplace-Lagrange expansion of the disturbing function is not guaranteed for high planetary eccentricities. Despite the high eccentricities of the planets, the expansion made of the secular terms only, when pushed to high order in eccentricities, can represent the orbits of non-resonant extrasolar systems with enough accuracy,
as shown by e.g. \cite{LibHen-2005} for the expansion at first order
in the masses and \cite{LibSan-2013} for the second order
expansion. However, in the Hamiltonians~(\ref{eq:HresO1}) and (\ref{eq:HresO2}), additional resonant terms are present and could restrain the
convergence domain.

To check the validity of the expansion for the 8 extrasolar systems
considered here, we have computed the boundaries of the domain of
convergence of the Laplace-Lagrange expansion of the disturbing
function, as given by the Sundman's criterion (\cite{Sun-1916}, see
also \cite{FM-1994}). The criterion for the absolute convergence of
the planar Laplace-Lagrange expansion is
\begin{equation}
 a_1F_1(e_1)<a_2F_0(e_2)
\end{equation}
with real functions $F_1(g)=\sqrt{1+g^2}\cosh{w}+g+\sinh{w}$ and $F_0(g)=\sqrt{1+g^2}\cosh{w}-g-\sinh{w}$, where $w=g \cosh{w}$. 

The boundaries of the convergence domain in $(e_1,e_2)$ space given by
the Sundman's criterion are shown in Fig.~\ref{figsundman} for several
semi-major axes ratios. The selected extrasolar systems are indicated
on the graph with the value of their semi-major axes ratio in
parenthesis. The convergence domain of each system is the domain below
the curve corresponding to the semi-major axes ratio of the system. We
observe that the eccentricities of HD~60532 and HD~108874 systems are
located well inside the convergence domain, while GJ~876 and GJ~3293
systems are slightly above the boundary curve. On the contrary, the
four remaining systems, HD~128311, HD~73526, HD~82943 and HD~45364
have eccentricities too large to be located inside their Sundman's
convergence domain, and the use of the expansion is not
appropriate. In the following, we will then focus on the long-term
evolutions of the four systems fulfilling the Sundman's
criterion. Their physical and orbital parameters are given in
Table~\ref{tab:param}. The last column of Table~\ref{tab:param} gives
an indication of the proximity of the system to the mean-motion
resonance (see Sect.~\ref{ss:delta}). The value of the
$\delta$-parameter clearly shows that GJ~876 and HD~60532 systems are
in the third category of systems, asking for a resonant approach,
while HD~108874 and GJ~3293 systems are only near the 4:1 resonance,
being close to the upper limit of the second regime.  Although a
secular approach at order two in the masses already gives a good
approximation of the dynamical evolution for these last two systems,
we aim to see whether the resonant approach yields to further
improvements.

\subsection{Validation of the resonant approximations}\label{sbs:an_int}
To assess the validity of the resonant Hamiltonians in describing the
long-term evolution of the planetary orbits, we will compare the
Runge-Kutta integration of the equations of motion associated to the
Hamiltonians~\eqref{eq:HresO1} and~\eqref{eq:HresO2} with the direct
numerical integration of the full problem (adopting the
$\Sscr\Bscr\Ascr\Bscr3$ symplectic scheme, see~\cite{LasRob-2001}).

We introduce the compact notations $\Cscr^{(\Oscr 1)}$ and
$\Cscr^{(\Oscr 2)}$ for the composition of the canonical
transformations defined in Section~\ref{sec:expansion}, namely
\begin{equation}
  \Cscr^{(\Oscr 1)}=\Tscr_{F}\circ\Tscr_{\Rscr}
  \quad\textrm{and}\quad
\Cscr^{(\Oscr 2)}=\Tscr_{F}\circ\Tscr_{\Oscr 2}\circ\Tscr_{\Rscr}\ .
\label{def-Cscr}
\end{equation}
Given initial values of the orbital elements
$\big(\avec(0),\lambdavec(0),\evec(0),\omegavec(0)\big)$, we compute
their evolution by exploiting the resonant Hamiltonian at order two in
the masses as follows
\begin{equation}
\vcenter{\openup1\jot\halign{
 \hbox to 25 ex{\hfil $\displaystyle {#}$\hfil}
&\hbox to 20 ex{\hfil $\displaystyle {#}$\hfil}
&\hbox to 25 ex{\hfil $\displaystyle {#}$\hfil}\cr
\big(\avec(0),\lambdavec(0),\evec(0),\omegavec(0)\big)
&\build{\xrightarrow{\hspace*{25pt}}}_{}^{{{\displaystyle
\left(\Cscr^{(\Oscr 2)}\right)^{-1}\circ\Escr^{-1}}
\atop \phantom{0}}}
&\left({{\displaystyle \Ivec(0)}
\,,\, {\displaystyle \sigmavec(0)}}\right)
\cr
& &
\bigg\downarrow 
\cr
\big(\avec(t),\lambdavec(t),\evec(t),\omegavec(t)\big)
&\build{\xleftarrow{\hspace*{25pt}}}_{}^{{{\displaystyle
\Escr\circ\Cscr^{(\Oscr 2)}} \atop \phantom{0}}}
&\left({{\displaystyle \Ivec(t)}
\,,\, {\displaystyle \sigmavec(t)}}\right)
\cr
}}
\ ,
\label{semi-analytical_scheme}
\end{equation}
where $(\Lambdavec,\lambdavec,\xivec,\etavec) =
\Escr^{-1}(\avec(0),\lambdavec(0),\evec(0),\omegavec(0))$ is the
non-canonical change of coordinates~\eqref{eq:poincvar}.  Of course,
the same scheme holds for the resonant Hamiltonian at order one in the masses with the change of $\Cscr^{(\Oscr 1)}$ in place of $\Cscr^{(\Oscr 2)}$.

In the following sections we detail the comparison of the long-term
evolutions of the eccentricities and possibly resonant angles given by
our analytical approach and by direct numerical integration for each
of the four considered extrasolar systems.

\begin{figure}
  \begin{center}
    \includegraphics[angle=270,width=0.95\textwidth]{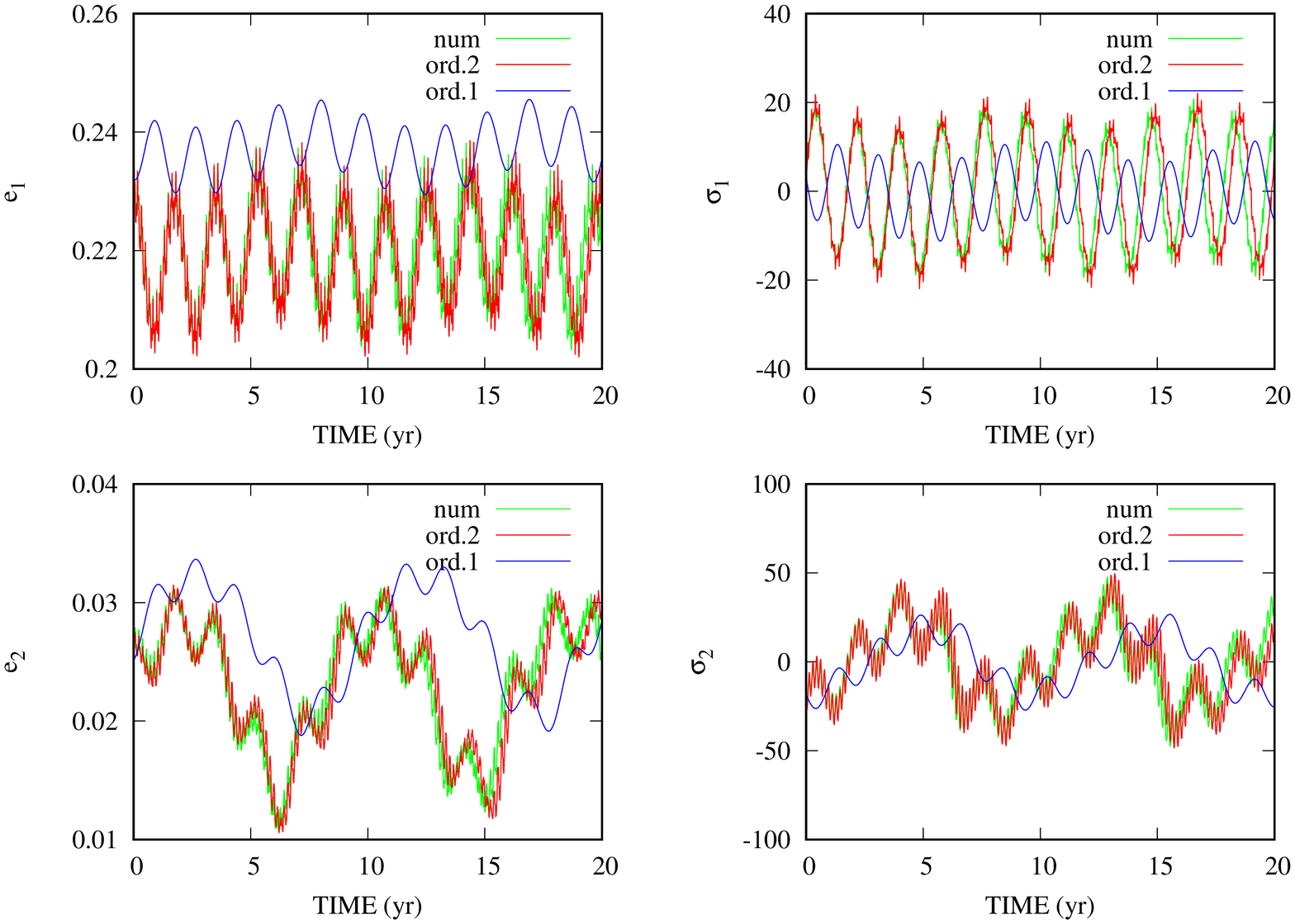}
  \end{center}
  \caption{Evolution of the eccentricities (left panels) and resonant angles (right panels) for GJ~876 system, given by (i)~direct
numerical integration (green curves); (ii)~second order resonant approximation (red curves); (iii)~first order resonant approximation (blue curves).}
 \label{figevolGJ876}

 \begin{center}
    \includegraphics[width=0.49\textwidth]{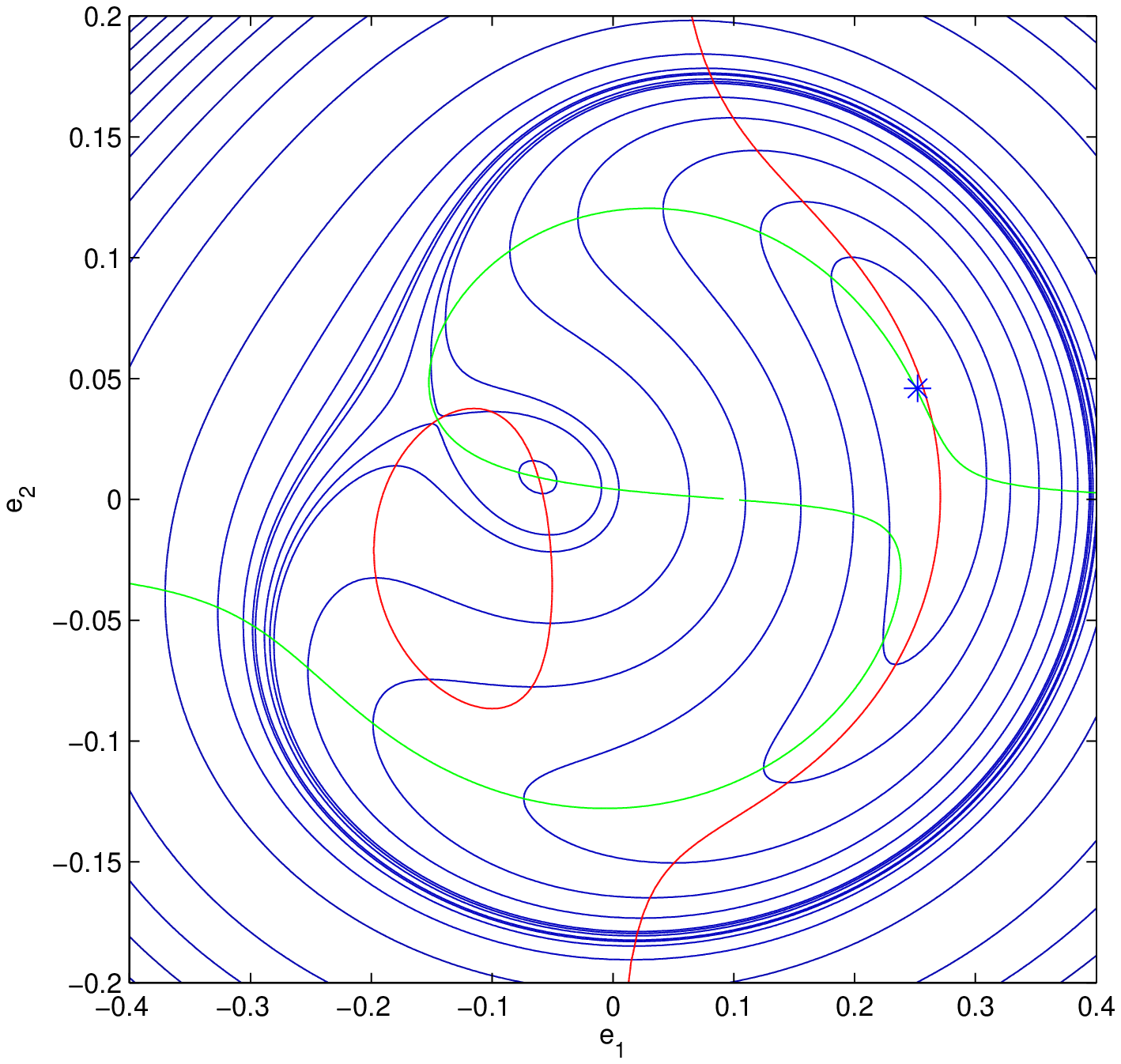}
    \includegraphics[width=0.49\textwidth]{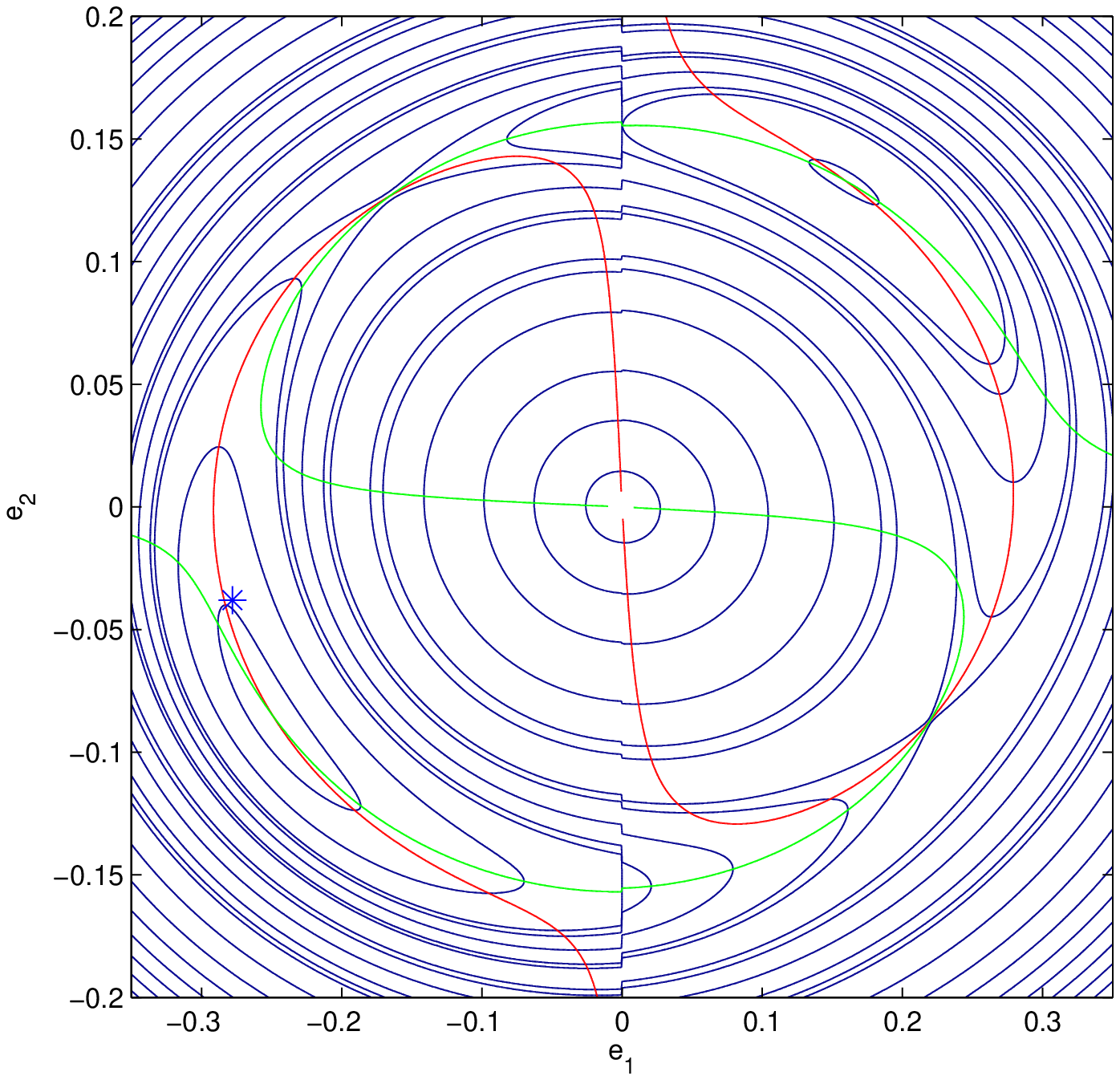}
  \end{center}
  \caption{Level curves of the first-order resonant Hamiltonian in the plane $(e_1 \sin{\sigma_1},e_2 \sin{\sigma_2})$ for GJ~876 (left panel) and in $(e_1 \sin{2\sigma_1},e_2 \sin{\Delta\omega})$  for HD~60532 (right panel), with the angles fixed to $0^\circ$ in the positive part of the axis and $180^{\circ}$ in the negative part of the axis. The stars indicate the locations of the extrasolar systems. See text for more details.}
 \label{figportrait}
\end{figure}

\begin{figure}
  \begin{center}
    \includegraphics[angle=270,width=0.8\textwidth]{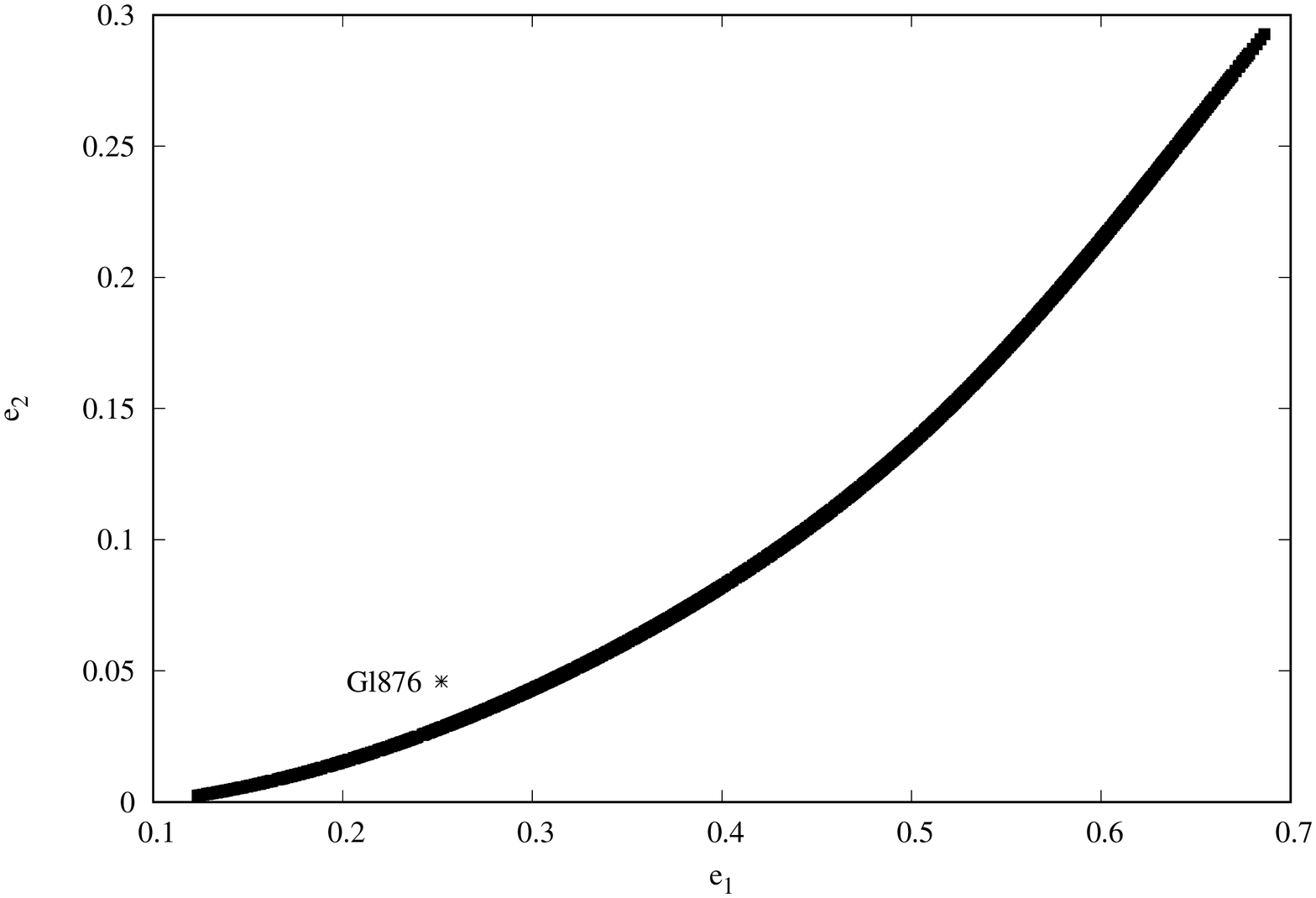}
  \end{center}
  \caption{Family of periodic orbits of the 2:1 mean-motion resonance corresponding to $(\sigma_1,\sigma_2)=(0,0)$ in the $(e_1,e_2)$ plane for the mass ratio of GJ~876, obtained with the first-order resonant approximation. The star indicates the eccentricities of GJ~876 system, demonstrating its proximity to the family of periodic orbits.}
 \label{figperiodic}
\end{figure}

\subsection{GJ~876}\label{sbs:GJ876}

One of the first detected resonant extrasolar systems was GJ~876
system locked in a 2:1 mean-motion resonance
(\cite{Mar-2001,Lau-2001}). The dynamical evolution of the two-planet
system was carefully studied in~\cite{Bea-2003}, where they used an
analytical expansion specifically devised for high-eccentricity
planetary three-body problem. Moreover, \cite{Ver-2007} showed the limitations of a resonant Hamiltonian expanded up to order
four in the eccentricities. Here we aim to perform a similar study of
GJ~876 system using the resonant Hamiltonian at order two in the
masses\footnote{Let us note that \cite{Riv-2010} have revealed the
presence of an additional planet in a three-body Laplace resonance with the
previously two known giant planets.}.

In Fig.~\ref{figevolGJ876}, we show the evolution of the
eccentricities (left panels) and the resonant angles (right panels) of
the two-planet (c-b pair) GJ~876 system, as given by our analytical approach. The parameters $K_S$ and $K_F$ have been fixed so as
to include the effects up to the third resonant harmonics, namely
$K_S=3$ and $K_F=9$. The blue curves
indicate the orbital evolutions obtained with the Hamiltonian
expansion at first order in the masses, while the evolution with the
second-order expansion is given in red. Fig.~\ref{figevolGJ876} shows that both resonant angles of
GJ~876 planetary system librate, confirming that the system is locked
in a 2:1 mean-motion resonance.

The accuracy of our
Hamiltonian approximation is verified by comparison with a direct
numerical integration of the Newton equations (green
curves). We observe in Fig.~\ref{figevolGJ876} that the resonant Hamiltonian to first order in the masses, although giving a rather good approximation of the
frequencies of the motion, does not reproduce reliably the variation
amplitudes of the eccentricities and the resonant angles. On the
contrary, the second-order approximation is very efficient and the
evolutions given by the analytical approach and the numerical
integration do superimpose nearly perfectly.

To analyze more deeply the dynamics of the two-planet GJ~876 system,
we reproduce in Fig.~\ref{figportrait} (left panel) the level curves
of the first-order Hamiltonian in the representative plane $(e_1
\sin{\sigma_1},e_2 \sin{\sigma_2})$, where both resonant angles are
fixed to $0^\circ$ in the positive part of the axis and $180^\circ$ in
the negative part of the axis, as previously done by
\cite{Bea-2003}. We insist that this plane is neither a phase
portrait, nor a surface of section, since the problem is
four-dimensional. To find the stationary solutions of the (averaged)
Hamiltonian, we have computed the curves $\dot{\sigma_1}=0$ and
$\dot{\sigma_2}=0$ on the representative plane (green and red curves),
since their intersections give the eccentricities of the stationary
solutions. The star symbol in Fig.~\ref{figportrait} indicates the
location of GJ~876 planetary system. We observe that the system lies
close to a stationary solution (also often denoted ACR, for apsidal
corotation resonance, see e.g. \cite{Bea-2003}), thus in a region
where both resonant angles librate.

The stationary solutions of the Hamiltonian averaged over the short
periods correspond to periodic orbits of the full problem. Using our
first-order approach, we have computed, in Fig.~\ref{figperiodic}, the
family of periodic orbits corresponding to $(\sigma_1,\sigma_2)=(0,0)$
in the $(e_1,e_2)$ plane, for the mass ratio of GJ~876. Comparisons
can be made with the works of \cite{Had-2002} and \cite{Bea-2003},
showing the accuracy of the analytical approximation. As expected, we observe
the proximity of GJ~876 system to the family of periodic orbits.

It is interesting to note that the approximation to first order in the
masses is accurate enough to compute both the dynamics on the
representative plane and the family of periodic orbits, giving a good
qualitative representation of the dynamics. As a result, we can
conclude that, while the resonant Hamiltonian at first order in the
masses gives a good indication on the resonant dynamics of GJ~876
system, the second-order correction is needed to reproduce carefully
the orbital evolution of the system.

\subsection{HD~60532}\label{sbs:HD60532}

\begin{figure}
  \begin{center}
    \includegraphics[angle=270,width=0.95\textwidth]{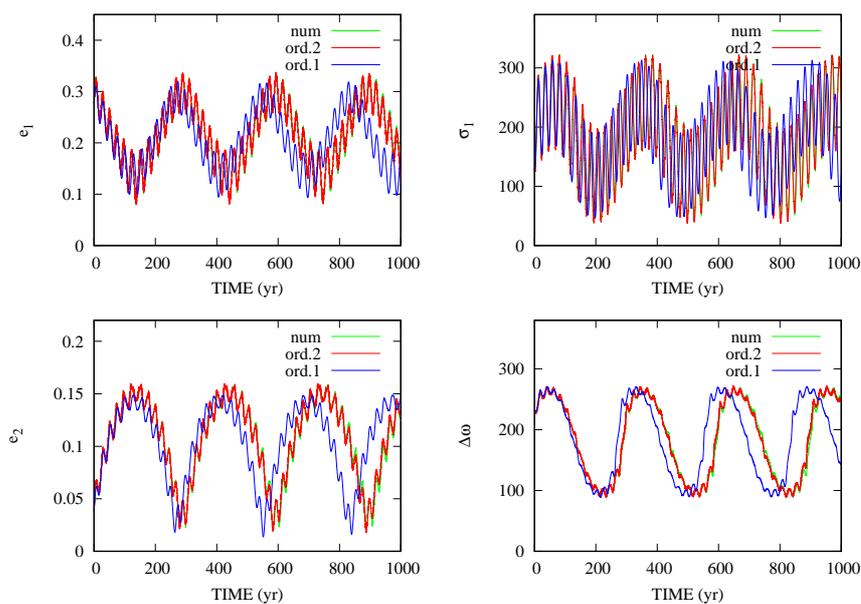}
  \end{center}
  \caption{Same as Fig.~\ref{figevolGJ876} for HD~60532 system.}
 \label{figevolHD60532}
\end{figure}

HD~60532 system consists of two
giant planets in a 3:1 mean-motion resonance (\cite{Des-2008}).  Two exhaustive
dynamical analyses of the system have been performed
by~\cite{Laskar-2009} and~\cite{Alv-2015}. Likewise the previous study, the evolutions of the eccentricities and
resonant angles are reported in Fig.~\ref{figevolHD60532}.  The
parameters $K_S$ and $K_F$ have been fixed so as to include the
effects up to the second resonant harmonics, namely $K_S=4$ and
$K_F=8$.  Fig.~\ref{figevolHD60532} shows that both the resonant angle
$\sigma_1$ and the difference of the longitudes~$\Delta\omega$
librate, confirming that the system is locked in a 3:1 mean-motion
resonance. This can also be deduced from Fig.~\ref{figportrait} (right panel), which shows the level curves of the resonant Hamiltonian at first order in the masses, in the representative plane $(e_1
\sin{2\sigma_1},e_2 \sin{\Delta \omega})$, where both resonant angles
are fixed to $0^\circ$ in the positive part of the axis and
$180^\circ$ in the negative part of the axis. Again, a perfect agreement with the results presented in \cite{Alv-2015} is observed.

The same conclusions can be drawn on the accuracy of the analytical approach, as for GJ~876 system. The resonant Hamiltonian at order one
in the masses gives already a good approximation of the long-term evolutions in  
Fig.~\ref{figevolHD60532}, while the one at order two reliably reproduces  the frequencies of the motion and the variation amplitudes of the orbital elements.

\subsection{HD~108874}\label{sbs:HD108874}

\begin{figure}
  \begin{center}
    \includegraphics[angle=270,width=0.95\textwidth]{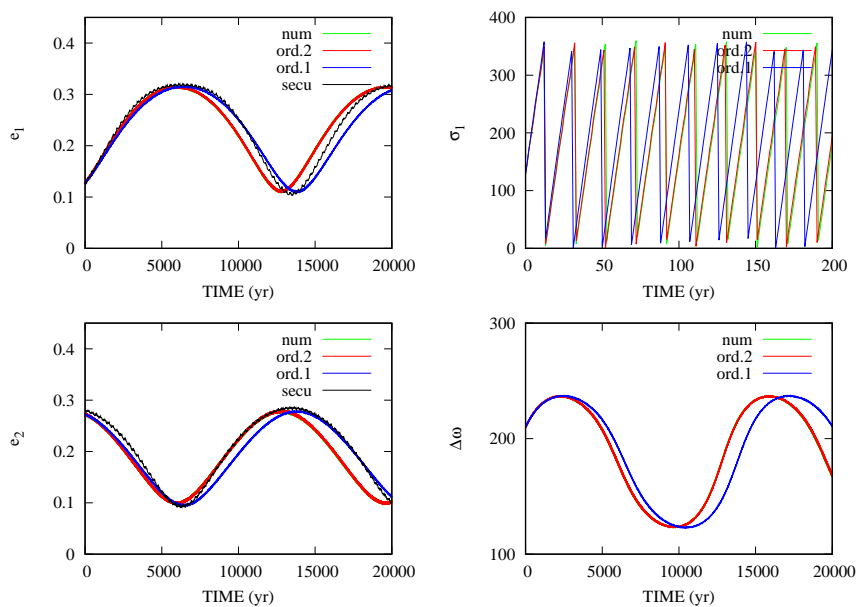}
  \end{center}
  \caption{Same as Fig.~\ref{figevolGJ876} for HD~108874 system.}
 \label{figevolHD108874}
\end{figure}

\begin{figure}
  \begin{center}
    \includegraphics[angle=270,width=0.95\textwidth]{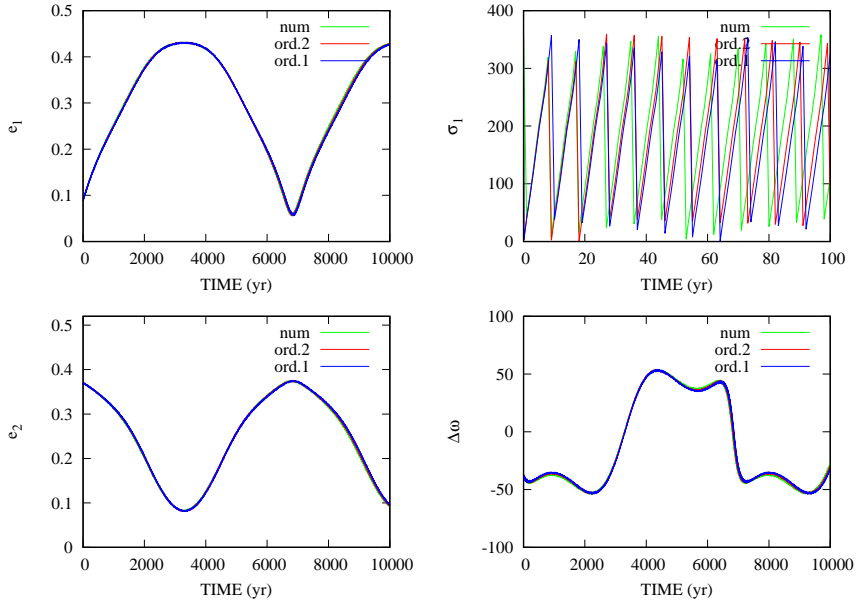}
  \end{center}
  \caption{Same as Fig.~\ref{figevolGJ876} for GJ~3293 system.}
 \label{figevolGJ3293}
\end{figure}

HD~108874 system (e.g.,~\cite{Butler-2003,Vogt-2005,WriUpaMarFis-2009}) is composed of two giant planets very close to the 4:1 mean-motion resonance, as previously deduced from the $\delta$-parameter value of the system (see the discussion at the beginning of Section~\ref{sec:appl}). Secular Hamiltonian expansions both at first and second orders in the masses were considered in our previous contribution (\cite{LibSan-2013}), showing a good agreement for the amplitudes of the long-term evolutions of the eccentricities, but not for the secular frequencies of the motion (see the black curves in Fig.~\ref{figevolHD108874}).

The results obtained with the resonant Hamiltonian formulations are
reported in Fig.~\ref{figevolHD108874}. The parameters $K_S$ and $K_F$
have been fixed so as to include the effects up to the second resonant
harmonics, namely $K_S=6$ and $K_F=10$. The resonant angles circulate,
thus HD~108874 planetary system is not locked in a 4:1 mean-motion
resonance\footnote{Let us stress that, to better visualize the
  evolution of $\sigma_1$, we plot the evolution on a much smaller
  timescale.}. The difference of the longitudes~$\Delta\omega$
librates. We see that a resonant Hamiltonian approach is needed in
order to accurately reproduce the long-term dynamics of the system. In
particular, the resonant approximation at order two in the masses is
the most efficient, and the evolutions given by this analytical
approach superimpose with the direct numerical integration. As a
result, we conclude that, for systems very close to a mean-motion
resonance, the resonant approach is required to reproduce perfectly
the secular frequencies of the motion.

\subsection{GJ~3293}\label{sbs:GJ3293}
\cite{Ast-2015} reported the detection of two-Neptune-like planets
around GJ~3293 with periods near the 4:1 commensurability. Let us note that
the parameters of GJ~3293 $b$ and $c$ have slightly changed with the
discovery of two additional Super-Earths in the system
(\cite{Ast-2017}), however the dynamical analysis of the four-body system
is beyond the scope of the present work.

GJ~3293 system behaves similarly to HD~108874, also being very close
to the same mean-motion resonance.  However, the smaller
  value of the $\delta$-parameter indicates that the system is less
  close to the resonance.  As a result, a resonant Hamiltonian at
order one in the masses already describes accurately the long-term
evolutions of the orbits, as illustrated in
Fig.~\ref{figevolGJ3293}. Still, looking more closely to the orbital
evolutions, we can appreciate the improvements due to the second order
approximation.

\section{Conclusions}\label{sec:ccl}
In the present work we have analyzed the long-term evolution of four
exoplanetary systems by using two different approximations: the
resonant Hamiltonians at order one and two in the masses.  Both
approximations include the (near-)resonant harmonics, extending the
classical Laplace-Lagrange secular approximation adopted to describe
the long-term evolution of non-resonant systems. This allows us to
treat the systems which are really close to or in a mean-motion
resonance, and for which the secular approximation failed in
~\cite{LibSan-2013}. Let us note that we only consider here planetary
systems for which the Sundman's criterion is fulfilled. We assess the
proximity of a system to the resonance by using the
$\delta$-parameter. This criterion reveals accurate, and confirms the
need of a resonant Hamiltonian at order two in the masses for high
values of the $\delta$-parameter ($\delta>2.6\times10^{-2}$), while a
resonant Hamiltonian at order one in the masses gives a good
approximation of the long-term evolution of the system for
$2.0\times10^{-2}<\delta<2.6\times10^{-2}$.

We considered two systems, HD~108874 and GJ~3293, which are {\it really close to} a 4:1 mean-motion resonance, but not in the resonance.  For
these systems the resonant Hamiltonian at order one already gives a
good approximation, while the resonant Hamiltonian at order two
perfectly represents the long-time evolution of these systems.

The second order approximation, due to the careful treatment of the
contribution of low-order fast harmonics, gives a substantial impact
when dealing with {\it resonant} systems.  This was illustrated in the
study of GJ~876 and HD~60532 systems, where the difference between resonant approximations at order one and two in the masses is much more noticeable.

\begin{acknowledgements}
 The authors seize the opportunity of the Topical Collection for the
 50th birthday of CM\&DA to dedicate this paper to the memory of
 Jacques Henrard. This work follows the path traced in his two
 contributions published in the first volume of Celestial Mechanics.
 The work of M.~S.  has been partially supported by the National Group
 of Mathematical Physics (GNFM-INdAM).  Computational resources have
 been provided by the PTCI (Consortium des \'Equipements de Calcul
 Intensif CECI), funded by the FNRS-FRFC, the Walloon Region, and the
 University of Namur (Conventions No. 2.5020.11, GEQ U.G006.15,
 1610468 et RW/GEQ2016).

\end{acknowledgements}

\bigskip

\emph{The authors have no conflict of interest to declare.}

\appendix

\section{Low-order expansions of $\Hscr_{{\rm res}}^{(\Oscr1)}$ and $\Hscr_{{\rm res}}^{(\Oscr2)}$ for GJ~876}\label{sec:A_exp}

We report here the low order expansion of $\Hscr_{{\rm res}}^{(\Oscr
  1)} = \overline\Hscr^{(\Tscr)} + \widetilde\Hscr^{(\Tscr)}$ and
$\Hscr_{{\rm res}}^{(\Oscr 2)} = \overline\Hscr^{(\Oscr 2)} +
\widetilde\Hscr^{(\Oscr 2)}$ (see~\eqref{eq:HresO1}
and~~\eqref{eq:HresO2}, respectively) for GJ~876. We refer to 
Table~\ref{tab:param} for the physical and orbital
parameters of the system and Subsection~\ref{sbs:GJ876} for a complete description
of the system.

\renewcommand{\arraystretch}{2.}
\begin{table}[H]
\caption{Low-order secular contributions of the resonant Hamiltonians
  at order one and two in the masses, namely
  $\overline\Hscr^{(\Tscr)}$ and $\overline\Hscr^{(\Oscr 2)}$.}
\label{tab:O1-O2}
\begin{center}
  \begin{tabular}{|c|c|r l|c|c|c|c|r|}
    \hline
    $L_1$ & $L_2$ & $ {\sin\atop\cos}(\lambda_1$ & $\lambda_2$) &
    $\xi_1$ & $\xi_2$ & $\eta_1$ & $\eta_2$ & Coefficient ${{\rm order\ } 1\atop {\rm order\ } 2}$ \\
    \hline
    0&    0&   $ \cos(0$&  $  0)$&    0&    0&    0&    0&  $-3.3464020777229581e+00\atop-3.3464563762212642e+00$\\
    \hline
    0&    0&   $ \cos(0$&  $  0)$&    2&    0&    0&    0&  $-1.0357954358101849e-01\atop-1.5743993289438937e-01$\\
    0&    0&   $ \cos(0$&  $  0)$&    1&    1&    0&    0&  $ 7.7180750141755586e-02\atop1.5099105592771100e-01$\\
    0&    0&   $ \cos(0$&  $  0)$&    0&    0&    2&    0&  $-1.0357954358101849e-01\atop-1.5743993289438937e-01$\\
    0&    0&   $ \cos(0$&  $  0)$&    0&    0&    1&    1&  $ 7.7180750141755586e-02\atop1.5099105592771100e-01$\\
    0&    0&   $ \cos(0$&  $  0)$&    0&    2&    0&    0&  $-2.6285032427253777e-02\atop-5.1442194373599197e-02$\\
    0&    0&   $ \cos(0$&  $  0)$&    0&    0&    0&    2&  $-2.6285032427253777e-02\atop-5.1442194373599197e-02$\\
    \hline
    1&    0&   $ \cos(0$&  $  0)$&    0&    0&    0&    0&  $ 7.5441751449930875e+01\atop7.5437939181811231e+01$\\
    0&    1&   $ \cos(0$&  $  0)$&    0&    0&    0&    0&  $ 3.7825543474951097e+01\atop3.7826365596719974e+01$\\
    \hline    	      	   	     	       	   
    1&    0&   $ \cos(0$&  $  0)$&    2&    0&    0&    0&  $-2.7620769522807734e+01\atop-5.1880019095053086e+01$\\
    0&    1&   $ \cos(0$&  $  0)$&    2&    0&    0&    0&  $ 9.6542831681610277e+00\atop1.5932534382562174e+01$\\
    1&    0&   $ \cos(0$&  $  0)$&    1&    1&    0&    0&  $ 2.6339667486758934e+01\atop5.7583867417714245e+01$\\
    0&    1&   $ \cos(0$&  $  0)$&    1&    1&    0&    0&  $-8.6550508609898422e+00\atop-1.6429487543265010e+01$\\
    1&    0&   $ \cos(0$&  $  0)$&    0&    0&    2&    0&  $-2.7620769522807734e+01\atop-5.1880019095053086e+01$\\
    0&    1&   $ \cos(0$&  $  0)$&    0&    0&    2&    0&  $ 9.6542831681610277e+00\atop1.5932534382562174e+01$\\
    1&    0&   $ \cos(0$&  $  0)$&    0&    0&    1&    1&  $ 2.6339667486758934e+01\atop5.7583867417714245e+01$\\
    0&    1&   $ \cos(0$&  $  0)$&    0&    0&    1&    1&  $-8.6550508609898422e+00\atop-1.6429487543265010e+01$\\
    1&    0&   $ \cos(0$&  $  0)$&    0&    2&    0&    0&  $-7.8909142498375440e+00\atop-1.8090303237027724e+01$\\
    0&    1&   $ \cos(0$&  $  0)$&    0&    2&    0&    0&  $ 2.6736770713809226e+00\atop5.1709156508143646e+00$\\
    1&    0&   $ \cos(0$&  $  0)$&    0&    0&    0&    2&  $-7.8909142498375440e+00\atop-1.8090303237027724e+01$\\
    0&    1&   $ \cos(0$&  $  0)$&    0&    0&    0&    2&  $ 2.6736770713809226e+00\atop5.1709156508143646e+00$\\
    \hline
  \end{tabular}
\end{center}
\end{table}

\renewcommand{\arraystretch}{2.}
\begin{table}[H]
\caption{Low-order resonant contributions of the resonant Hamiltonians
  at order one and two in the masses, namely
  $\widetilde\Hscr^{(\Tscr)}$ and $\widetilde\Hscr^{(\Oscr 2)}$.}
\label{tab:O1-O2res}
\begin{center}
  \begin{tabular}{|c|c|l r|c|c|c|c|r|}
    \hline
    $L_1$ & $L_2$ & $ {\sin\atop\cos}(\lambda_1$ & $\lambda_2$) &
    $\xi_1$ & $\xi_2$ & $\eta_1$ & $\eta_2$ & Coefficient ${{\rm order\ } 1\atop {\rm order\ } 2}$ \\
    \hline
    0&    0&   $\cos(\phantom{-} 1$&  $ -2)$&    1&    0&    0&    0&  $ 5.5376484645283615e-02\atop5.7238823791435821e-02$\\
    0&    0&   $\sin(-1$&  $  2)$&    0&    0&    1&    0&  $-5.5376484645283615e-02\atop-5.7238823791435821e-02$\\
    0&    0&   $\cos(\phantom{-} 1$&  $ -2)$&    0&    1&    0&    0&  $-9.8782747719408249e-03\atop-1.1153891516908212e-02$\\
    0&    0&   $\sin(-1$&  $  2)$&    0&    0&    0&    1&  $ 9.8782747719408249e-03\atop1.1153891516908212e-02$\\
    \hline
    0&    0&   $\cos(\phantom{-} 2$&  $ -4)$&    2&    0&    0&    0&  $-4.5217344035489809e-01\atop-4.8651258965168032e-01$\\
    0&    0&   $\sin(-2$&  $  4)$&    1&    0&    1&    0&  $ 9.0434688070979619e-01\atop9.7302517930336063e-01$\\
    0&    0&   $\cos(\phantom{-} 2$&  $ -4)$&    1&    1&    0&    0&  $ 6.7028607445890176e-01\atop7.2398694788154001e-01$\\
    0&    0&   $\sin(-2$&  $  4)$&    1&    0&    0&    1&  $-6.7028607445890176e-01\atop-7.2398695137503255e-01$\\
    0&    0&   $\cos(\phantom{-} 2$&  $ -4)$&    0&    0&    2&    0&  $ 4.5217344035489809e-01\atop4.8651258965168032e-01$\\
    0&    0&   $\sin(-2$&  $  4)$&    0&    1&    1&    0&  $-6.7028607445890176e-01\atop-7.2398695137503255e-01$\\
    0&    0&   $\cos(\phantom{-} 2$&  $ -4)$&    0&    0&    1&    1&  $-6.7028607445890176e-01\atop-7.2398694788154001e-01$\\
    0&    0&   $\cos(\phantom{-} 2$&  $ -4)$&    0&    2&    0&    0&  $-2.4544612025031667e-01\atop-2.6718668544376162e-01$\\
    0&    0&   $\sin(-2$&  $  4)$&    0&    1&    0&    1&  $ 4.9089224050063335e-01\atop5.3437364957117151e-01$\\
    0&    0&   $\cos(\phantom{-} 2$&  $ -4)$&    0&    0&    0&    2&  $ 2.4544612025031667e-01\atop2.6718668544376162e-01$\\
    \hline
    1&    0&   $\cos(\phantom{-} 1$&  $ -2)$&    1&    0&    0&    0&  $ 9.3088616131547308e+00\atop9.3325713373268862e+00$\\
    0&    1&   $\cos(\phantom{-} 1$&  $ -2)$&    1&    0&    0&    0&  $-3.5407106603304044e+00\atop-3.5483602184707737e+00$\\
    1&    0&   $\sin(-1$&  $  2)$&    0&    0&    1&    0&  $-9.3088616131547308e+00\atop-9.3325713373268862e+00$\\
    0&    1&   $\sin(-1$&  $  2)$&    0&    0&    1&    0&  $ 3.5407106603304044e+00\atop3.5483602184707737e+00$\\
    1&    0&   $\cos(\phantom{-} 1$&  $ -2)$&    0&    1&    0&    0&  $-5.7283159393729672e+00\atop-5.7368751996508980e+00$\\
    0&    1&   $\cos(\phantom{-} 1$&  $ -2)$&    0&    1&    0&    0&  $ 1.6638688261793673e+00\atop1.6659529830384545e+00$\\
    1&    0&   $\sin(-1$&  $  2)$&    0&    0&    0&    1&  $ 5.7283159393729672e+00\atop5.7368751996508980e+00$\\
    0&    1&   $\sin(-1$&  $  2)$&    0&    0&    0&    1&  $-1.6638688261793673e+00\atop-1.6659529830384545e+00$\\
    1&    0&   $\cos(\phantom{-} 2$&  $ -4)$&    2&    0&    0&    0&  $-1.3443610498475394e+02\atop-1.4382396459577518e+02$\\
    0&    1&   $\cos(\phantom{-} 2$&  $ -4)$&    2&    0&    0&    0&  $ 4.5662300692010547e+01\atop4.7951527384253403e+01$\\
    1&    0&   $\sin(-2$&  $  4)$&    1&    0&    1&    0&  $ 2.6887220996950788e+02\atop2.8764825455632791e+02$\\
    0&    1&   $\sin(-2$&  $  4)$&    1&    0&    1&    0&  $-9.1324601384021094e+01\atop-9.5902937120616485e+01$\\
    \hline
  \end{tabular}
\end{center}
\end{table}

\renewcommand{\arraystretch}{2.}
\begin{center}
  \begin{tabular}{|c|c|l r|c|c|c|c|r|}
    \hline
    $L_1$ & $L_2$ & $ {\sin\atop\cos}(\lambda_1$ & $\lambda_2$) &
    $\xi_1$ & $\xi_2$ & $\eta_1$ & $\eta_2$ & Coefficient ${{\rm order\ } 1\atop {\rm order\ } 2}$ \\
    \hline
    1&    0&   $\cos(\phantom{-} 2$&  $ -4)$&    1&    1&    0&    0&  $ 1.6623038850909461e+02\atop 1.7920631597850823e+02$\\
    0&    1&   $\cos(\phantom{-} 2$&  $ -4)$&    1&    1&    0&    0&  $-5.9300450505275727e+01\atop-6.2736552403780614e+01$\\
    1&    0&   $\sin(-2$&  $  4)$&    1&    0&    0&    1&  $-1.6623038850909461e+02\atop-1.7920631008500678e+02$\\
    0&    1&   $\sin(-2$&  $  4)$&    1&    0&    0&    1&  $ 5.9300450505275727e+01\atop 6.2736534976756516e+01$\\
    1&    0&   $\cos(\phantom{-} 2$&  $ -4)$&    0&    0&    2&    0&  $ 1.3443610498475394e+02\atop 1.4382396459577515e+02$\\
    0&    1&   $\cos(\phantom{-} 2$&  $ -4)$&    0&    0&    2&    0&  $-4.5662300692010547e+01\atop-4.7951527384253410e+01$\\
    1&    0&   $\sin(-2$&  $  4)$&    0&    1&    1&    0&  $-1.6623038850909461e+02\atop-1.7920631008500683e+02$\\
    0&    1&   $\sin(-2$&  $  4)$&    0&    1&    1&    0&  $ 5.9300450505275727e+01\atop 6.2736534976756509e+01$\\
    1&    0&   $\cos(\phantom{-} 2$&  $ -4)$&    0&    0&    1&    1&  $-1.6623038850909461e+02\atop-1.7920631597850829e+02$\\
    0&    1&   $\cos(\phantom{-} 2$&  $ -4)$&    0&    0&    1&    1&  $ 5.9300450505275727e+01\atop 6.2736552403780600e+01$\\
    1&    0&   $\cos(\phantom{-} 2$&  $ -4)$&    0&    2&    0&    0&  $-4.8820026145600778e+01\atop-5.3339208015605962e+01$\\
    0&    1&   $\cos(\phantom{-} 2$&  $ -4)$&    0&    2&    0&    0&  $ 1.8656715340550328e+01\atop 1.9991782463540289e+01$\\
    1&    0&   $\sin(-2$&  $  4)$&    0&    1&    0&    1&  $ 9.7640052291201556e+01\atop 1.0667882985908244e+02$\\
    0&    1&   $\sin(-2$&  $  4)$&    0&    1&    0&    1&  $-3.7313430681100655e+01\atop-3.9983785581166615e+01$\\
    1&    0&   $\cos(\phantom{-} 2$&  $ -4)$&    0&    0&    0&    2&  $ 4.8820026145600778e+01\atop 5.3339208015605955e+01$\\
    0&    1&   $\cos(\phantom{-} 2$&  $ -4)$&    0&    0&    0&    2&  $-1.8656715340550328e+01\atop-1.9991782463540289e+01$\\
    \hline
  \end{tabular}
\end{center}


\begin{thebibliography}{50}
\makeatletter
\def\@biblabel#1{\hbox to 10pt{\hss[#1]}}
\makeatother

\bibitem[Alves et al.(2015)]{Alv-2015}{A. Alves, T. Michtchenko and M. Tadeu dos Santos: \emph{Dynamics of the 3/1 planetary mean-motion resonance. An application to the HD60532 b-c planetary system}, CeMDA, {\bf 124}, 311--334 (2015).}
  
\bibitem[Astudillo-Defru et al.(2015)]{Ast-2015}{N. Astudillo-Defru, X. Bonfils, X. Delfosse et al.: \emph{The HARPS search for southern extra-solar planets XXXV. Planetary systems and stellar activity of the M dwarfs GJ 3293, GJ 3341, and GJ 3543}, A\&A, {\bf 575}, A119, 19pp. (2015).}

\bibitem[Astudillo-Defru et al.(2017)]{Ast-2017}{N. Astudillo-Defru, T. Forveille, X. Bonfils et al.: \emph{The HARPS search for southern extra-solar planets XLI. A dozen planets around the M dwarfs GJ 3138, GJ 3323, GJ 273, GJ 628, and GJ 3293}, A\&A, {\bf 602}, A88, 21pp. (2017).}
  
\bibitem[Batygin \& Morbidelli (2013)]{Bat-2013}{K. Batygin and A. Morbidelli: \emph{Analytical treatment of planetary resonances}, A\&A, {\bf 556}, A28, 20pp (2013).}

\bibitem[Beaug\'e \& Michtchenko(2003)]{Bea-2003}{C. Beaug\'e and T. Michtchenko: \emph{Modelling the high-eccentricity planetary three-body problem. Application to the GJ876 planetary system.},
  MNRAS, {\bf 341}, 760 (2003).}

\bibitem[Butler et al.(2003)]{Butler-2003}{R.P. Butler, G.W. Marcy, S.S. Vogt, D.A. Fischer, G.W. Henry, G. Laughlin, J.T. Wright: \emph{Seven New Keck Planets Orbiting G and K Dwarfs}, The Astrophysical Journal, {\bf 582}, 455--466 (2003).}

\bibitem[Callegari et al.(2004)]{Cal-2004}{Callegari N. Jr., Michtchenko T.A. and Ferraz-Mello S.: \emph{Dynamics of two planets in the 2/1 mean-motion resonance}, CeMDA, {\bf 556}, 89, 201--234 (2004).}

\bibitem[Callegari et al.(2006)]{Cal-2006}{Callegari N. Jr., Ferraz-Mello S. and Michtchenko T.A.: \emph{Dynamics of Two Planets in the 3/2 Mean-motion Resonance: Application to the Planetary System of the Pulsar PSR B1257+12}, CeMDA, {\bf 94}, 381--397 (2006).}

\bibitem[Celletti \& Chierchia(2005)]{CelChi-2005}{A. Celletti and
  L. Chierchia: \emph{ KAM stability and celestial mechanics},
  Mem. Amer. Math. Soc., {\bf 187}, 1--134 (2007).}

\bibitem[Correia et al.(2009)]{Cor-2009}{A.C.M. Correia, S. Udry, M. Mayor et al.: \emph{The HARPS search for southern extra-solar planets - XVI. HD45364, a pair of planets in a 3:2 mean motion resonance},
  A\&A, {\bf 496}, 521--526 (2009).}
 
\bibitem[Desort et al.(2008)]{Des-2008}{M. Desort, A.-M. Lagrange,
  F. Galland, H. Beust, S. Udry, M. Mayor and G. Lo Curto:
  \emph{Extrasolar planets and brown dwarfs around A-F type
    stars. V. A planetary system found with HARPS around the F6IV-V
    star HD 60532 }, Astronomy and Astrophysics, {\bf 491}, 883--888 (2008).}

\bibitem[Duriez(1989a)]{Duriez-1989a}{L. Duriez: \emph{Le probl\`eme
    des deux corps revisit\'e}, Les M\'ethodes modernes de la
  M\'ecanique C\'eleste, Editions Fronti\`eres, 9--34 (1989a).}

\bibitem[Duriez(1989b)]{Duriez-1989b}{L. Duriez: \emph{Le
    d\'eveloppement de la fonction perturbatrice}, Les M\'ethodes
  modernes de la M\'ecanique C\'eleste, Editions Fronti\`eres, 35--62
  (1989b).}

\bibitem[Ferraz-Mello(1994)]{FM-1994}{S. Ferraz-Mello: \emph{The convergence domain of the Laplacian expansion of the disturbing function}, CeMDA, {\bf 58}, 37--52 (1994).}  
  
\bibitem[Gabern(2005)]{GabJorLoc-2005}{F. Gabern, A. Jorba,
  U. Locatelli: \emph{ On the construction of the Kolmogorov normal
    form for the Trojan asteroids}, Nonlinearity, {\bf 18} N.4,
  1705--1734 (2005).}

\bibitem[Giorgilli(1995)]{Giorgilli-1995}{A. Giorgilli:
  \emph{Quantitative methods in classical perturbation theory}, From
  Newton to chaos: modern techniques for understanding and coping with
  chaos in N­body dynamical systems, Nato ASI school, A.E. Roy e
  B.D. (1995).}

\bibitem[Giorgilli et al.(2009)]{GioLocSan-2009}{A. Giorgilli,
  U. Locatelli, M. Sansottera: \emph{Kolmogorov and Nekhoroshev theory
    for the problem of three bodies}, CeMDA, {\bf 104}, 159--173
  (2009).}

\bibitem[Giorgilli \& Sansottera(2011)]{GioSan-2012} {A. Giorgilli and
  M, Sansottera: \emph{ Methods of algebraic manipulation in
    perturbation theory}, Workshop Series of the Asociacion Argentina
  de Astronomia, {\bf 3}, 147--183 (2011).}

\bibitem[Giorgilli et al.(2017)]{GioLocSan-2017}{A. Giorgilli,
  U. Locatelli, M. Sansottera: \emph{Secular dynamics of a planar
    model of the Sun-Jupiter-Saturn-Uranus system; effective stability
    into the light of Kolmogorov and Nekhoroshev theories}, Regular
  and Chaotic Dynamics, {\bf 22}, 54--77 (2017).}

\bibitem[Hadjidemetriou(2002)]{Had-2002}{J. Hadjidemetriou: \emph{Resonant periodic motion and the stability of extrasolar planetary systems.},
  CeMDA, {\bf 83}, 141 (2002).}

\bibitem[Henrard(1973)]{Henrard-1973}{J. Henrard: \emph{The algorithm
    of the inverse for Lie transform}, Recent Advances in Dynamical
  Astronomy, Astrophysics and Space Science Library, {\bf 39},
  248--257 (1973).}

\bibitem[Laskar(1988)]{Laskar-1988}{J. Laskar: \emph{Secular evolution
    over 10 million years}, A\&A, {\bf 198}, 341--362 (1988).}

\bibitem[Laskar(1989)]{Laskar-1989}{J. Laskar: \emph{Syst\`emes de
    variables et \'el\'ements}, Les M\'ethodes modernes de la
  M\'ecanique C\'eleste, Editions Fronti\`eres, 63--87 (1989).}

\bibitem[Laskar \& Robutel(1995)]{LasRob-1995}{J. Laskar and
  P. Robutel: \emph{ Stability of the Planetary Three-Body Problem ---
    I. Expansion of the Planetary Hamiltonian}, CeMDA, {\bf 62},
  193-217 (1995).}

\bibitem[Laskar \& Robutel(2001)]{LasRob-2001}{J. Laskar and
  P. Robutel: \emph{High order symplectic integrators for perturbed
    Hamiltonian systems}, CeMDA, {\bf 80}, 39--62 (2001).}
  
\bibitem[Laskar \& Correia(2009)]{Laskar-2009}{J.Laskar and A.C.M. Correia: \emph{HD60532, a planetary system in a 3:1 mean motion resonance}, A\&A, {\bf 496}, L5 (2009).}
    
\bibitem[Laughlin \& Chambers(2001)]{Lau-2001}{G. Laughlin and J. Chambers: \emph{Short-term dynamical interactions among extrasolar planets}, ApJ, {\bf 551}, L109-L113 (2001).}

\bibitem[Libert \& Henrard(2005)]{LibHen-2005}{A.-S. Libert and
  J. Henrard: \emph{Analytical approach to the secular behaviour of
    exoplanetary systems}, CeMDA, {\bf 93}, 187--200 (2005).}

\bibitem[Libert \& Henrard(2007)]{LibHen-2007}{A.-S. Libert and
  J. Henrard: \emph{Analytical study of the proximity of exoplanetary
    systems to mean-motion resonances}, A\&A, {\bf 461}, 759--763
  (2007).}

\bibitem[Libert \& Sansottera(2013)]{LibSan-2013}{A.-S. Libert and
  M. Sansottera: \emph{On the extension of the Laplace-Lagrange secular theory
to order two in the masses for extrasolar systems}, 
  CeMDA, {\bf 117}, 149--168 (2013).}

\bibitem[Locatelli \& Giorgilli(2007)]{LocGio-2007}{U. Locatelli and
  A. Giorgilli: \emph{Invariant tori in the Sun–Jupiter–Saturn
    system}, DCDS-B, {\bf 7}, 377--398 (2007).}

\bibitem[Marcy et al.(2001)]{Mar-2001}{G. Marcy, P. Butler, D. Fisher et al.: \emph{A Pair of Resonant Planets Orbiting GJ 876.},
  ApJ, {\bf 556}, 296 (2001).}   

\bibitem[Poincar\'e(1893)]{Poi-1893}{H. Poincar\'e: \emph{Les
     m\'ethodes nouvelles de la M\'ecanique C\'eleste},
   Gauthier-Villars (1893).}

\bibitem[Rivera et al.(2010)]{Riv-2010}{E. Rivera, G. Laughlin, P. Butler et al.: \emph{The Lick-Carnegie Exoplanet Survey: A Uranus-mass Fourth Planet for GJ 876 in an Extrasolar Laplace Configuration},
 ApJ, {\bf 719}, 890 (2010).}
  
\bibitem[Robutel(1995)]{Robutel-1995}{P. Robutel: \emph{ Stability of
    the Planetary Three-Body Problem --- II. KAM Theory and Existence
    of Quasiperiodic Motions}, CeMDA, {\bf 62}, 219--261 (1995).}

\bibitem[Sansottera et al.(2011)]{SanLocGio-2011b}{M. Sansottera,
  U. Locatelli, A. Giorgilli: \emph{A Semi-Analytic Algorithm for
    Constructing Lower Dimensional Elliptic Tori in Planetary
    Systems}, CeMDA, {\bf 111}, 337--361 (2011).}

\bibitem[Sansottera et al.(2013)]{SanLocGio-2011a}{M. Sansottera,
  U. Locatelli, A. Giorgilli: \emph{On the stability of the secular
    evolution of the planar Sun-Jupiter-Saturn-Uranus system},
  Math. Comput. Simulat. {\bf 88}, 1--14 (2013).}

\bibitem[Sansottera et al.(2015)]{SanGraGio-2015}{M. Sansottera,
  L. Grassi, A. Giorgilli: \emph{On the relativistic Lagrange-Laplace
    secular dynamics for extrasolar systems}, Proceedings of the IAU
  Symposium S310, 74--77 (2015).}

\bibitem[Sundman(1916)]{Sun-1916}{K.F. Sundman: \emph{Sur les conditions nécessaires et suffisantes pour la convergence du développement de la fonction perturbatrice dans le mouvement plan}, \"Ofversigt Finska Vetenskaps-Soc. F\"orh {\bf 58A}, 24 (1916).}    
    
\bibitem[Tan et al.(2013)]{Tan-2013}{X. Tan, M. Payme, M.H. Lee et al.: \emph{Characterizing the orbital and dynamical state of the HD 82943 planetary system with Keck radial velocity data}, ApJ, {\bf 777}, id. 101, 21 pp. (2013).}
    
\bibitem[Veras(2007)]{Ver-2007}{D. Veras:\emph{A resonant-term-based model including a nascent disk, precession, and oblateness: application to GJ 876},
CeMDA, {\bf 99}, 197--243 (2007).}
  
\bibitem[Vogt et al.(2005)]{Vogt-2005}{S.S. Vogt, R.P. Butler, G.W. Marcy, D.A. Fischer, G.W. Henry, G. Laughlin, J.T. Wright, J.A. Johnson: \emph{Five New Multicomponent Planetary Systems}, The Astrophysical Journal, {\bf 632}, 638--658 (2005).}

\bibitem[Wittenmyer et al.(2014)]{Wit-2014}{R.A. Wittenmyer, X. Tan,
  M.H. Lee et al.: \emph{A detailed analysis of the HD 73526 2:1 resonant planetary system}, ApJ, {\bf 780}, id.140, 9pp. (2014).}    
  
\bibitem[Wright et al.(2009)]{WriUpaMarFis-2009}{J.T. Wright,
  S. Upadhyay, G.W. Marcy, D.A. Fisher et al.: \emph{Ten new and
    updated multiplanet systems and a survey of exoplanetary systems},
  ApJ, {\bf 693}, 1084--1099 (2009).}

  


  
\end{thebibliography}
\end{document}